\documentclass[runningheads]{llncs}
\usepackage{graphicx}
\usepackage{floatrow}
\usepackage{amsmath}
\usepackage{amssymb}
\def\*#1{\mathbf{#1}}
\def\*t#1{\mathbf{\tilde #1}}
\def\^#1{\boldsymbol{#1}}

\def\pr#1{\left( #1\right)}

\def\*#1{\mathbf{#1}}

\begin{document}
\title{Knitting a Markov blanket is hard when you are out-of-equilibrium: two examples in canonical nonequilibrium models}
\titlerunning{Knitting a Markov blanket is hard when you are out-of-equilibrium}
%
\author{Miguel Aguilera\inst{1}  
\and Ángel Poc-López\inst{2}
\and
Conor Heins\inst{3}
\and Christopher L. Buckley\inst{1}
}
\authorrunning{M. Aguilera et al.}
%
\institute{School of Engineering and Informatics, University of Sussex, Falmer, Brighton, United Kingdom\\ \email{sci@maguilera.net, C.L.Buckley@sussex.ac.uk} \and
ISAAC Lab, I3A Engineering Research Institute of Aragon, University of Zaragoza, Zaragoza, Spain. \\
\email{angel.poc.lopez@gmail.com} \and
Department of Collective Behaviour, Max Planck Institute of Animal Behavior, Konstanz, Germany\\
\email{cheins@ab.mpg.de}}

%
\maketitle              
\begin{abstract}
Bayesian theories of biological and brain function speculate that Markov blankets (a conditional independence separating a system from external states) play a key role for facilitating inference-like behaviour in living systems. Although it has been suggested that Markov blankets are commonplace in sparsely connected, nonequilibrium complex systems, this has not been studied in detail. Here, we show in two different examples (a pair of coupled Lorenz systems and a nonequilibrium Ising model) that sparse connectivity does not guarantee Markov blankets in the steady-state density of nonequilibrium systems. Conversely, in the nonequilibrium Ising model explored,  the more distant from equilibrium the system appears to be correlated with the distance from displaying a Markov blanket. These result suggests that further assumptions might be needed in order to assume the presence of Markov blankets in the kind of nonequilibrium processes describing the activity of living systems.

\keywords{Markov blankets \and Nonequilibrium dynamics \and Bayesian inference \and Lorenz attractor \and Ising model.}
\end{abstract}
\section{Introduction}

In statistical inference, a Markov blanket describes a
subset of variables containing all the required information to infer the state of another subset. Identifying a Markov blanket reduces the computational complexity of inferring generative models of some variables to capturing dependencies with blanket states.
Specifically, a Markov blanket describes a set of variables (the `blanket') separating two other sets of variables, that become independent conditioned on the state of the blanket. If a system $\*s=\{s_{1},s_2,\dots,s_N\}$
can be decomposed into three subsets $\*x$, $\*b$ and $\*y$, $\*b$ is a Markov blanket if it renders $\*x$, $\*y$ conditionally independent:
\begin{align}
	p(\*x,\*y|\*b) = p(\*x|\*b)p(\*y|\*b).
	\label{eq:Markov-blanket}
\end{align}
This property, also referred to as the global Markov condition \cite{richardson1996automated}, implies an absence of functional couplings between $\*x$ and $\*y$, given the blanket $\*b$. 

\begin{figure}
\floatbox[{\capbeside\thisfloatsetup{capbesideposition={right,top},capbesidewidth=5.2cm}}]{figure}[\FBwidth]
{\caption{Sparse structural connectivity. The FEP assumes that Markov blankets naturally arise (under some conditions) when internal and external states are not structurally connected \cite{friston2021some}. All the models explored in this article will display this sparse connectivity pattern.}\label{fig:connectivity}}
{\includegraphics[width=4.8cm]{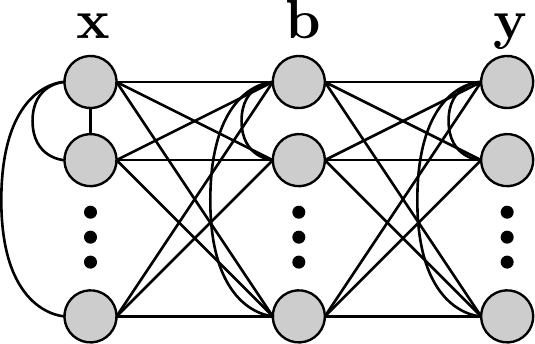}}
\end{figure}

Beyond its role as a technical tool for statistical inference, Markov blankets are becoming a subject of discussion in Bayesian approaches to biological systems, specially in literature addressing the free energy principle (FEP). The FEP is a framework originating in theoretical neuroscience promoting a Bayesian perspective of the dynamics of self-organizing adaptive systems (including living organisms) \cite{friston2013life,friston2019free,friston2022free}. The FEP claims that the internal states of certain systems can be described as if they were (approximately) inferring  the hidden sources of sensory variations. Its foundational literature  assumes that Markov blankets emerge from a sparse structural connectivity, decoupling internal states of a self-organizing system from its environmental milieu, (external states), via some interfacing states (Fig. \ref{fig:connectivity}) -- e.g., the cell's membrane, or a combination of retinal and oculomotor states during vision. The assumption is that this sparse connectivity leads to a statistical decoupling of internal states conditioned on the blanket \cite{friston2013life}. 
Although different versions of the theory address different aspects of the idea of a Markov blanket (e.g. its temporal evolution \cite{parr2021memory} or its role in paths outside a stationary density \cite{friston2022free}), in the present article, we restrict our analysis of Markov blankets to the `traditional' formulation of the FEP \cite{friston2019free}, where conditional independence relationships are expected to hold between states in the steady-state probability density that defines a stochastic system.  


\subsubsection{Acyclic networks.}

\begin{figure}
    \centering
    \includegraphics[width=8cm]{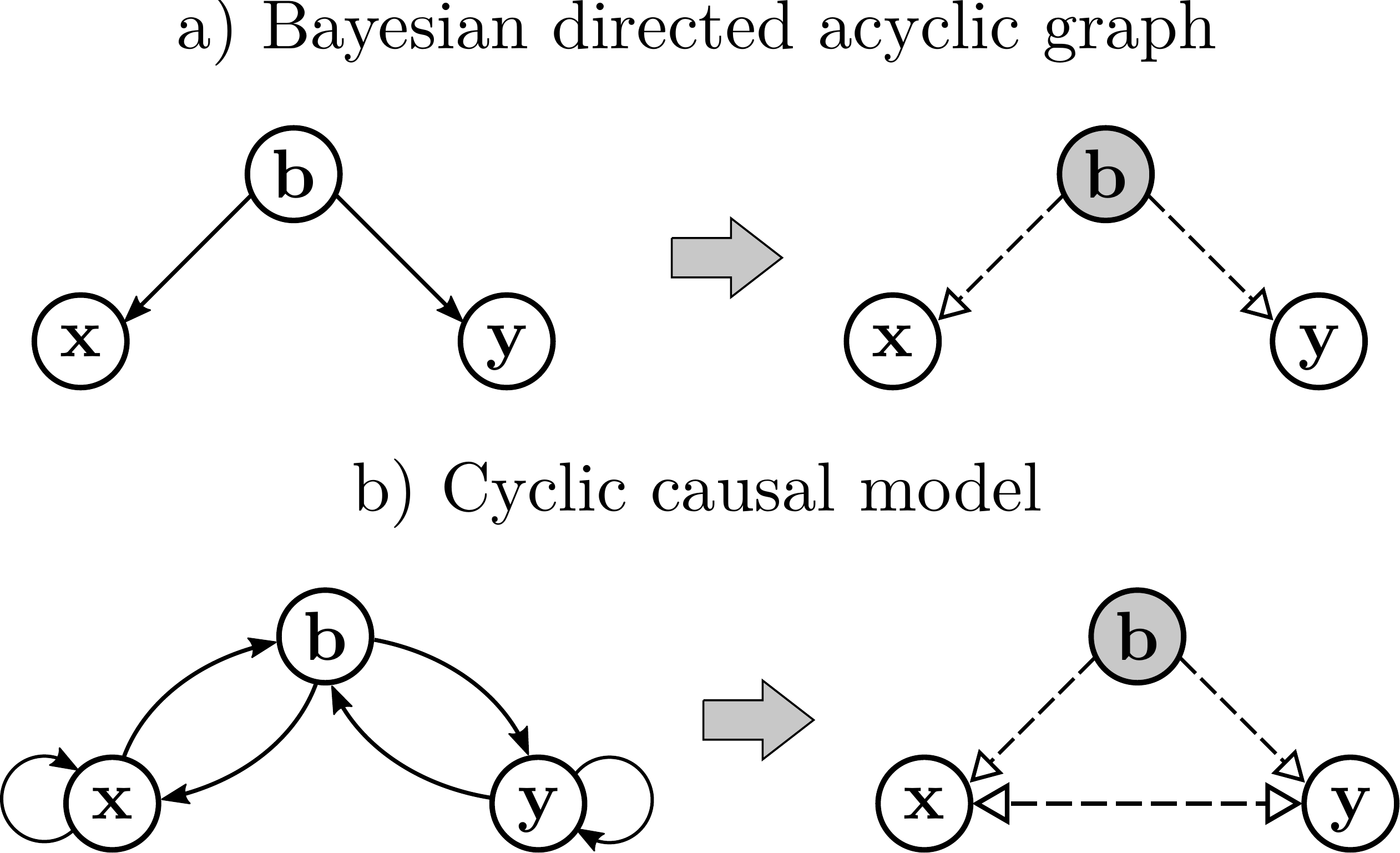}
    \caption{The left-hand figures show the structural connectivity of directed graphs. The right-hand figures show the conditional functional couplings of the system when the state of the `blanket' $\*b$ is fixed. In directed acyclic graphs (a), the structural and functional couplings are directly related, and fixing the boundary results in conditional independence of $\*x,\*y$, yielding a Markov blanket. In directed cyclic graphs (b), the recurrent structural connections result in additional functional couplings between variables, generating new couplings between $\*x, \*y$ that `cross' the boundary, therefore no longer rendering $\*b$ a Markov blanket in general.}
    \label{fig:DAGs}
\end{figure}

Originally, Markov blankets were introduced in acyclic Bayesian networks \cite{pearl1988probabilistic}, where they can be identified using a simple rule applied over the structural connections of the network (e.g. Fig. \ref{fig:DAGs}.A). By this rule, the Markov blanket $\*b$ of a subset $\*x$ contains the parent nodes of $\*x$, the children nodes of $\*x$ and the parents of each child.
This specific sparse structural connectivity is defined as the local Markov condition \cite{richardson1996automated}.
Originally, the FEP derived its intuitions about Markov blankets from acyclic models, considering the local Markov condition for a Markov blanket \cite{friston2013life,friston2019free}, suggesting that a boundary between system and environment arises naturally from this sparse structural connectivity as in directed acyclic graphs, without considering functional dynamics.

\subsubsection{Equilibrium systems.}

More recent literature on the FEP justifies a similar equivalence of Markov blankets and structural connectivity under an asymptotic approximation to a weak-coupling equilibrium \cite[see Eq. S8 in Supplementary Material]{friston2021parcels}. Under this assumption, it has been predicted that Markov blankets will be commonplace in adaptive systems, e.g., in brain networks \cite{hipolito2021markov,friston2021some}. 
It is easy to observe that many instances of equilibrium systems will display Markov blankets under sparse, pairwise connectivity (Fig. \ref{fig:connectivity}).
For example, consider any causal system described as a dynamical Markov chain in discrete time:
\begin{align}
    p(\*s_t) = \sum_{\*s_{t-1}}w(\*s_t|\*s_{t-1}) p(\*s_t|\*s_{t-1}),
\end{align}
or its continuous state-time equivalent using a master equation:
\begin{align}
    \frac{d p(\*s_t)}{d t} = \int_{\*s'_{t}}  w(\*s_t|\*s'_{t}) p(\*s'_{t})d\*s'_{t},
\end{align}
where $w$ describes transition probabilities between states. 
Eventually, if the system converges to a global attractor, it will be described by a probability distribution
\begin{align}
    p(\*s) = Z^{-1} \exp\pr{-\beta E(\*s)}
    \label{eq:eq_distribution}
\end{align}
where $Z$ is a partition function. In thermodynamic equilibrium $E(\*s)$ will capture the Hamiltonian function of the system.
Thermodynamic equilibrium implies a condition called `detailed balance', which requires that, in steady state, transitions are time-symmetric, i.e., $w(\*s|\*s')p(\*s')  = w(\*s'|\*s)p(\*s)$, resulting in
\begin{align}
     \frac{w(\*s|\*s')}{w(\*s'|\*s)} \propto \exp\pr{-\beta (E(\*s)-E(\*s')) }.
    \label{eq:detailed_balance}
\end{align}

If a system is described by the sparse connectivity structure in Fig. \ref{fig:connectivity}, then its energy can be decomposed into
\begin{align}
    E(\*s) = E_\mathrm{int}(\*x,\*b) + E_\mathrm{ext}(\*b,\*y),
\end{align}
leading to a conditional indpendence
\begin{align}
    p(\*x,\*y|\*b) =& Z_{\*b}^{-1} \exp\pr{ - \beta E_\mathrm{int}(\*x,\*b) } \cdot \exp\pr{ -\beta E_\mathrm{ext}(\*b,\*y)}
     = p(\*x|\*b) p(\*y|\*b) .
    \label{eq:functional-decoupling}
\end{align}

\subsubsection{Recurrent, nonequilibrium systems.}

The most recent arguments in favour of why sparse coupling implies conditional independence follow from analysis of a stochastic system's coupling structure using a Helmholtz decomposition re-describing a continuous Langevin equation in terms of a gradient flow on the system's (log) stationary probability \cite{tomita1974irreversible,graham1977covariant,eyink1996hydrodynamics,pavliotis2014stochastic}. Briefly,  a dynamical system  described by a (Ito) stochastic differential equation:
\begin{equation}
    \frac{d\*s_t}{dt} = f(\*s_t) + \varsigma(\*s_t) \^\omega 
\end{equation}
where $f$ is the drift or deterministic part of the flow, $\varsigma$ is the diffusive or stochastic part (which can be state-dependent) and $\^\omega$ a Wiener noise with covariance $2\Gamma(\*s_t)$. The  Helmholtz decomposition expresses $f$ as follows \cite[Equation 3]{heins2022sparse}:
\begin{align}
    f(\*s) &= \underbrace{- \Gamma(\*s) \nabla E(\*s) + \nabla \cdot \Gamma(\*s)}_{\text{dissipative}} + \underbrace{Q(\*s)\nabla E(\*s) - \nabla \cdot Q(\*s)}_{\text{solenoidal}},
    \label{eq:helmholtz}
\end{align}
expressing the total drift $f$ as a gradient flow on the log of the stationary density $E(\*s)\propto \log p(\*s)$.
This decomposition involves two orthogonal gradient fields, a dissipative (or curl-free) term and a rotational (or  divergence-free) term. 

In a system subject only to dissipative forces, Eq.~  \ref{eq:helmholtz} is compatible with Eq.~\ref{eq:detailed_balance}
for continuous-time systems. In contrast,  a system driven by nonequilibrium dynamics will no longer show a direct correspondence between its dynamics and  steady state distribution, thus a Markov blanket is not guaranteed from sparse connectivity.
Given this difficulty \cite{biehl2021technical,aguilera_how_2022}, recent extensions of the FEP require additional conditions besides the absence of solenoidal couplings $Q(\*s)$ between internal and external states to guarantee a Markov blanket \cite{friston2021some}.
Nevertheless, a recent exploration of nonequilibrum linear systems showed that these extra conditions are unlikely to emerge without stringent restrictions of the parameter space  \cite{aguilera_how_2022}.
In such linear systems, their cyclic, asymmetric structure propagates reverberant activity system-wide, generating couplings beyond their structural connectivity (e.g. Fig. \ref{fig:DAGs}.B). As a consequence, for most parameter configurations of a system, the sparse connectivity of the local Markov condition does not result in a Markov blanket. That is, even if a system only interacts with the environment via a physical boundary, it will in general not display the conditional independence associated with a Markov blanket \cite{aguilera_how_2022}.
Recently, these arguments have been dismissed under the argument that living systems are poorly described by linear dynamics and thermodynamic equilibrium, and thus the scope of the FEP is focused on non-equilibrium systems \cite{friston2022very}. Further work has argued Markov blankets may appear in high-dimensional state-spaces and spatially-localized interactions \cite{heins2022sparse}, under the assumption of a quadratic potential. The rest of this paper will explore how likely are Markov blankets to emerge for canonical nonlinear out-of-equilibrium models.

\section{Results}

To test empirically the extent to which Markov blankets can be expected out of equilibrium, we have performed conditional independence tests over two canonical non-linear systems: the Lorenz system and the asymmetric kinetic Ising model. Lorenz systems have long been studied due to their chaotic behaviour \cite{lorenz1963deterministic}. In contrast, asymmetric kinetic Ising models are recently becoming a popular tool to study non-equilibrium biological systems like neural networks \cite{roudi2015multi,aguilera_how_2022}.
\subsubsection{Measure of conditional independence.}
Markov blanket conditional independence (Eq. \ref{eq:Markov-blanket}) implies an absence of functional couplings between internal states $\*x$ and external states $\*y$ once the value of the blanket $\*b$ is fixed. This condition is captured by the conditional mutual information being equal to zero:
\begin{align}
    I(\*x;\*y|\*b) = \sum_{\*x,\*b,\*y} p(\*x,\*b,\*y) \log \frac{ p(\*x,\*y|\*b) }{p(\*x|\*b)p(\*y|\*b)}
    \label{eq:cond_MI}
\end{align}
This conditional mutual information is equivalent to the Kullback Leibler divergence $D_\mathrm{KL}(p(\*x,\*y|\*b)||p(\*x|\*b)p(\*y|\*b))$, i.e. the dissimilarity between the joint and conditionally independent probability distributions. Thus, it is trivial to show that Eq. \ref{eq:Markov-blanket} holds only and only if  $I(\*x;\*y|\*b) =0$.

\subsubsection{Pair of coupled Lorenz systems.}

\begin{figure}
\floatbox[{\capbeside\thisfloatsetup{capbesideposition={right,top},capbesidewidth=4.2cm}}]{figure}[\FBwidth]
{\caption{Pair of coupled Lorenz systems. a) Conditional mutual information $I(\*x;\*y|\*b)$ of the coupled (solid line, $\chi=0.5$) and decoupled (dashed line, $\chi=0$) system, estimating using a 25 bin 6-dimensional histogram. b) Comparison of the joint and independent probability densities (estimated for a 100 bin bidimensional histogram) of variables $x_2,y_2$.}\label{fig:lorenz}}
{    \begin{tabular}{c}
        \multicolumn{1}{l}{a)} \\
            \includegraphics[width=6cm]{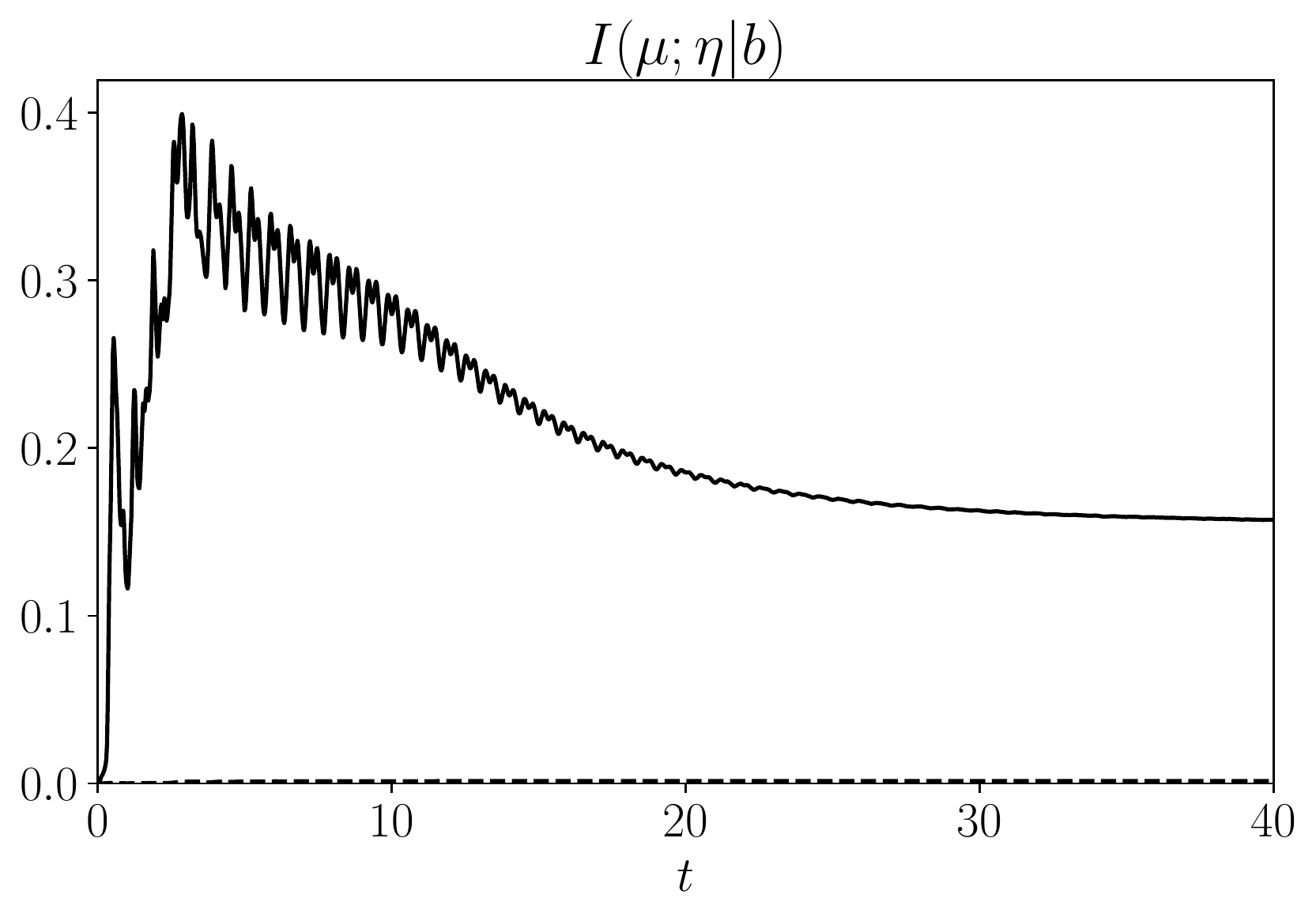}
        \\
        \multicolumn{1}{l}{b)}  \\
        \includegraphics[width=6cm]{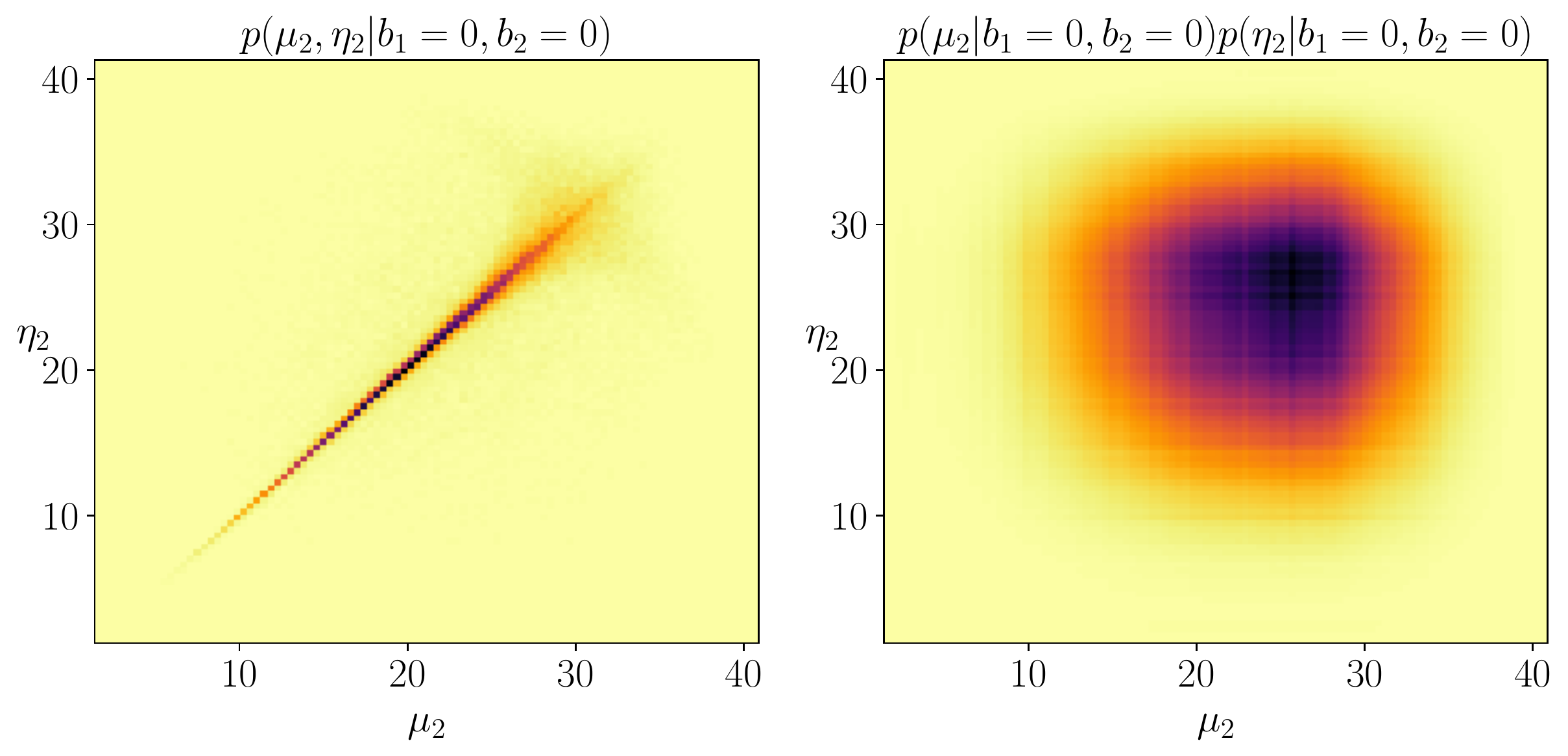}
    \end{tabular}}
\end{figure}

In \cite{friston2021stochastic}, the authors explore a system composed of two coupled Lorenz systems. The Lorenz system is a three-dimensional system of differential equations first studied by Edward Lorenz \cite{lorenz1963deterministic}, displaying chaotic dynamics for certain parameter configurations. The system explored in \cite{friston2021stochastic} describes two three-dimensional systems that are coupled to each other through the states $b_1$ and $b_2$. The equations of motion for the full six-dimensional system are: 
\begin{align}
    \frac{d}{d t} 
\begin{pmatrix}
        b_{1,t} \\  x_{1,t} \\ x_{2,t}\\ b_{2,t} \\ y_{1,t} \\ y_{2,t}
\end{pmatrix}= 
\begin{pmatrix}
        \sigma (x_{1,t} - \chi b_{2,t} - (1-\chi) b_{1,t} )
        \\ \rho b_{1,t} - x_{1,t}  - b_{1,t}x_{2,t}
        \\ b_{1,t}x_{1,t} -  \beta x_{2,t}
        \\ \sigma (y_{1,t} -  \chi b_{1,t} - (1-\chi) b_{2,t})
        \\ \rho b_{2,t} - y_{1,t}  - b_{2,t}y_{2,t}
        \\ b_{2,t}y_{1,t} -  \beta y_{2,t}
\end{pmatrix}
\end{align}
with  $\sigma = 10$, $\beta = 8/3$, and $\rho = 32$. The coupling parameter is set to $\chi=0.5$ (we will use  $\chi=0$ as reference of an uncoupled system) expecting the system to display nonequilibrium, chaotic dynamics.
Even in the absence of random fluctuations, the chaotic nature of the system will result in a rich steady-state probability distribution $p(\*s_t)$.
In \cite{friston2021stochastic}, authors show a Markov blanket conditional independence (Eq. \ref{eq:Markov-blanket}) by approximating $p(\*s_t)$ with a multivariate Gaussian (the so-called `Laplace assumption' \cite{parr2021memory}).
A careful analysis of the conditional mutual information $I(\*x;\*y|\*b)$ reveals that the system does not display a Markov blanket. In Fig.~\ref{fig:lorenz}.a we show the conditional mutual information $I(\*x;\*y|\*b)$ of the coupled Lorenz systems (solid line, $\chi=0.5$), estimating over an ensemble of $10^7$ trajectories from a random starting point (each variable $\mathcal{N}(0,1)$), and estimating its probability density using a histogram with 25 bins for each of the 6 dimensions. In comparison, the pair of decoupled Lorenz systems (dashed line, $\chi=0$), shows near zero conditional mutual information only due to sampling noise). We note that the authors of \cite{friston2021stochastic} never claim that the true stochastic Lorenz system (or the coupled equivalent) has Markov blankets, only that their Laplace-approximated equivalents do.


\subsubsection{Nonequilibrium kinetic Ising model.}

The asymmetric kinetic Ising model is a dynamical model with asymmetric couplings between binary spins $\*s$ (with values $\pm1$) at times $t$ and $t-1$, describing spin updates as:
\begin{align}
	w(s_{i,t}|\*s_{t-1}) &= \frac{\exp\pr{ s_{i,t} h_{i,t}}}{2 \cosh h_{i,t} }
	\\ h_{i,t} &= \sum_j J_{ij} s_{j,t-1}
\end{align}
We define asynchronous dynamics in which, at each time step, only one spin is updated.
In the case of symmetric couplings, $J_{ij}=J_{ji}$, the system converges to an equilibrium steady state, guaranteed by the detailed balance condition $p(\*s)$ maximum entropy distribution (Eq.~\ref{eq:eq_distribution}), with $E(\*s)=\sum_{ij} J_{ij} s_i s_j$, displaying emerging phenomena like critical phase transitions maximizing information integration and transfer \cite{aguilera2019integrated}.
In the case of asymmetric couplings, the system converges to a nonequilibrium steady state distribution $p(\*{s}_{t})$, generally displaying a complex statistical structure with higher-order interactions \cite{aguilera2021unifying}. In contrast with static equilibrium systems, asymmetries in $\*J$ result in loops of oscillatory activity involving a nonequilibrium entropy production \cite{aguilera2022nonequilibrium}, corresponding to entropy dissipation in a steady-state irreversible process. In stochastic thermodynamics this is described as the divergence between forward and reverse trajectories \cite{jarzynskiHamiltonianDerivationDetailed2000,crooks1998nonequilibrium}, relating the system’s time asymmetry with the entropy change of the reservoir. 
The entropy production $\sigma_t$ at time $t$ is then given as
\begin{align}
	\sigma_{t} 
	=& \sum_{\*s_t,\*s_{t-1}}p(\*s_t,\*s_{t-1})
	\log \frac{w(\*s_t|\*s_{t-1}) p(\*s_{t-1})}{w(\*s_{t-1}|\*s_{t})p(\*s_{t})},
	\label{eq:entropy_production}
\end{align}
which is the Kullback-Leibler divergence between the forward and backward trajectories \cite{schnakenberg_network_1976,seifert_stochastic_2012,itoUnifiedFrameworkEntropy2020}.

\begin{figure}
    \begin{center}
    \begin{tabular}{ll}
       \includegraphics[width=4.5cm]{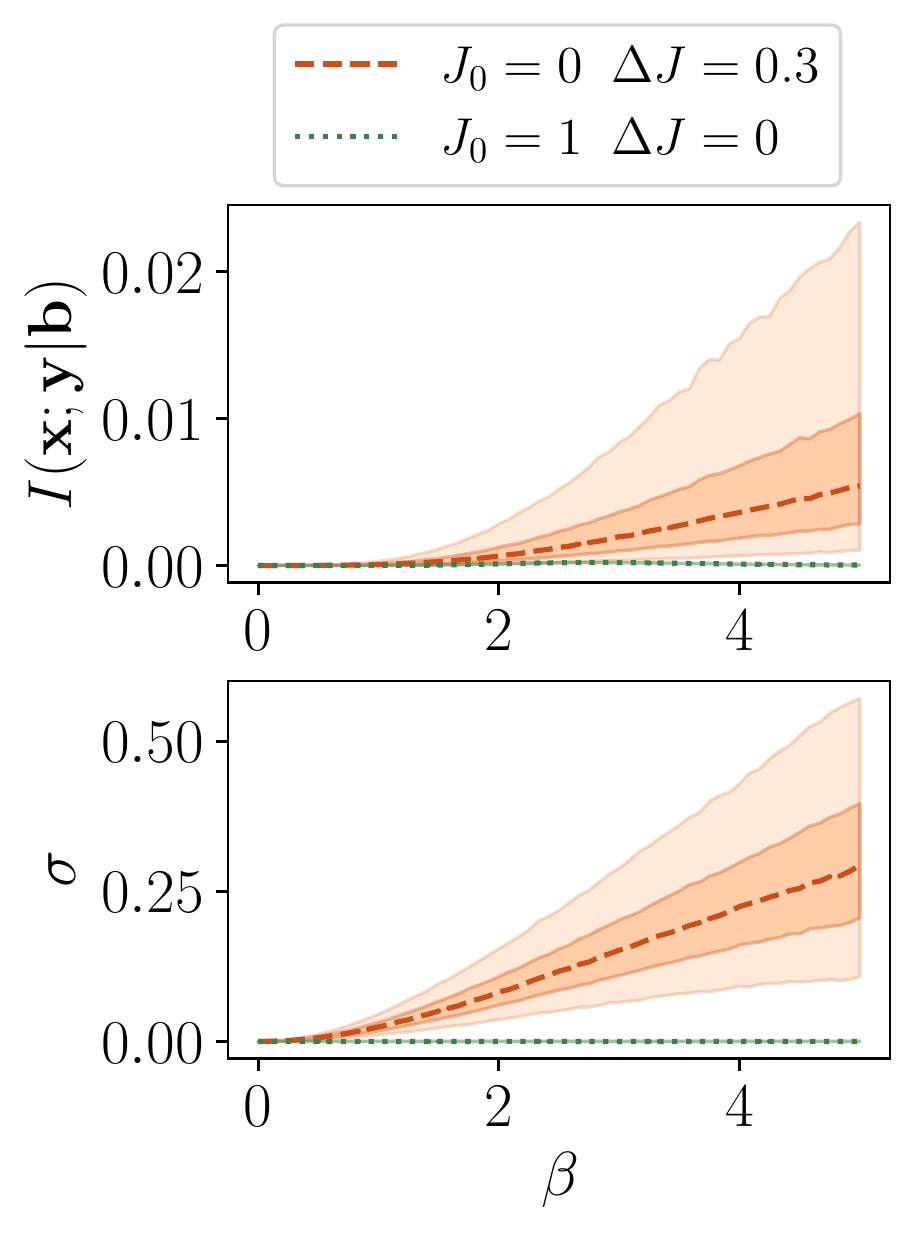} & \includegraphics[width=4.5cm]{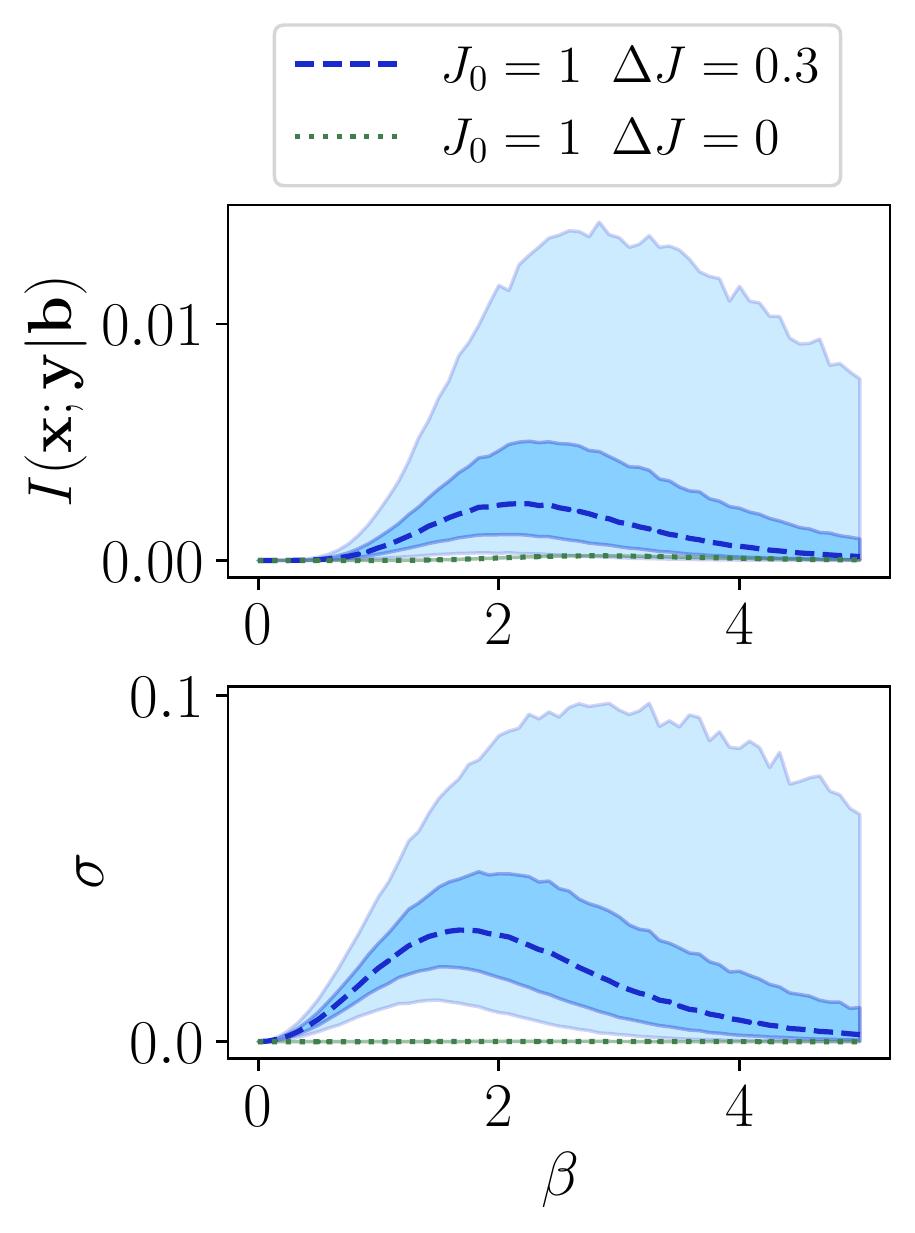}  
    \end{tabular}
    \end{center}
    
    \caption{Conditional mutual information $I(\*x;\*y|\*b)$ (top) and entropy production $\sigma$ (bottom) at different inverse temperatures ($\beta$) for a kinetic Ising model with Gaussian couplings and connectivity as in Fig.~\ref{fig:connectivity}, including systems with disordered dynamics ($J_0=0,\Delta J = 0.3$, red curves), a nonequilibrium order-disorder transition ($J_0=1,\Delta J = 0.3$, blue curves), and an equilibrium transition ($J_0=1,\Delta J = 0$, green curves). Areas show the median,  25/75 and 5/95 percentiles  for $10^4$ configurations.}
    \label{fig:ising}
\end{figure}

In the asymmetric Ising model, when couplings $J_{ij}$ have a Gaussian distribution $\mathcal{N}(J_0/N,\Delta J^2/N)$ (an asymmetric equivalent of the Sherrington-Kirkpatrick model). In the thermodynamic limit the system generates out-of-equilibrium structures both in an order-disorder critical phase transition ($\Delta J < \Delta J_c$), and in a regime showing highly-deterministic disordered dynamics ($\Delta J > \Delta J_c$ and large $\beta$) \cite{aguilera2022nonequilibrium,aguilera2021unifying}. Here, we will study in detail a network with just 6 nodes and random couplings with the connectivity in Fig.~\ref{fig:connectivity} where the probability distribution $p(\*s_t)$ can be calculated exactly. We will use parameters corresponding to an order-disorder phase transition ($J_0=1,\Delta J = 0.3$) and a disordered dynamics ($J_0=0,\Delta J = 0.3)$). We will compare the results with the behaviour of the system in equilibrium, when disorder between couplings is removed ($J_0=1,\Delta J = 0$).
The equilibrium system (equivalent to independent functional couplings as in Eq.~\ref{eq:functional-decoupling}), results in a Markov blanket with zero conditional mutual information $I(\*x;\*y|\*b)$, as well as zero entropy production $\sigma$ (Fig.~\ref{fig:ising}, red line). Nonetheless, this is not the case when couplings are asymmetric (Fig.~\ref{fig:ising}). Out of equilibrium, we observe how as the entropy production increases (i.e., the further the system is from equilibrium), the larger is the  conditional mutual information $I(\*x;\*y|\*b)$ (i.e., the further the system is from displaying a Markov blanket). This is particularly noticeable around the transition point in the order-disorder transition ($J_0=1,\Delta J = 0.3$), suggesting that Markov blankets might be specially challenging near nonequilibrium critical points.

\subsubsection{Discussion.}
These results raise fundamental concerns about the frequent use of Markov blankets as an explanatory concept in studying the behaviour of biological systems.
Our results however suggest that additional assumptions are needed for Markov blankets to arise under nonequilibrium conditions.
In consequence, without further assumptions, it may not be possible to take for granted that biological systems operate in a regime where Markov blankets arise naturally.
We shall note that the examples explored here have a reduced dimensionality (6 variables for both the Lorenz and asymmetric Ising systems). Previous work in the literature has suggested that high-dimensionality might be required to guarantee Markov blankets \cite{friston2021stochastic}, \cite{heins2022sparse}, but this remains a speculation. Further work could extend the type of analysis performed here to larger-dimensional systems.


%
%
\bibliographystyle{splncs04}
\bibliography{references}




\end{document}